\documentclass[conference]{IEEEtran}

\usepackage{graphicx,xcolor}
\usepackage[fleqn]{amsmath}

\usepackage[percent]{overpic}

\usepackage{amsthm}
\usepackage{amssymb}
\usepackage{amsfonts}

\usepackage{newtxtext}
\usepackage[varg]{newtxmath}
\usepackage{here}

\usepackage{mathtools}
\mathtoolsset{showonlyrefs=true}

\let\labelindent\relax
\usepackage[inline, shortlabels]{enumitem}
\usepackage{tablefootnote}
\usepackage{mathdots}
\usepackage{algorithmic}
\usepackage{amssymb}
\usepackage{algorithm}
\usepackage{xspace}

\usepackage{mathrsfs}

\makeatletter
\newcommand\RedeclareMathOperator{%
  \@ifstar{\def\rmo@s{m}\rmo@redeclare}{\def\rmo@s{o}\rmo@redeclare}%
}
\newcommand\rmo@redeclare[2]{%
  \begingroup \escapechar\m@ne\xdef\@gtempa{{\string#1}}\endgroup
  \expandafter\@ifundefined\@gtempa
     {\@latex@error{\noexpand#1undefined}\@ehc}%
     \relax
  \expandafter\rmo@declmathop\rmo@s{#1}{#2}}
\newcommand\rmo@declmathop[3]{%
  \DeclareRobustCommand{#2}{\qopname\newmcodes@#1{#3}}%
}
\@onlypreamble\RedeclareMathOperator
\makeatother

\definecolor{grey}{rgb}{0.7, 0.75, 0.71}

\def\dark-red#1{\textcolor[rgb]{0.7,0.0,0.0}{#1}}
\def\black#1{\textcolor[rgb]{0.0,0.0,0.0}{#1}}
\definecolor{amber}{rgb}{1.0, 0.75, 0.0}

\def\...{\dotsc}

\def\intT2{\int_{-T/2}^{T/2}}

\def\sumi1n{\sum_{i=1}^{n}}
\def\sumi1N{\sum_{i=1}^{N}}
\def\sumi0N--{\sum_{i=0}^{N-1}}

\def\ccc{\cdots}

\def\=def{\overset{\text{\small def}}{=}}
\newcommand{\defeq}{\overset{\text{\small def}}{=}}

\newcommand{\argmin}{\operatornamewithlimits{argmin}}

\DeclarePairedDelimiterX{\inp}[2]{\langle}{\rangle}{#1, #2}

\def\Cc{\mathcal{C}}

\def\Ic{\mathcal{I}}

\def\Sc{\mathcal{S}}

\def\<{\langle}
\def\>{\rangle}

 \def\mat4#1#2#3#4{
\begin{pmatrix}
 #1&\ccc&#2\\
 \vdots&&\vdots\\
 #3&\ccc&#4
\end{pmatrix}}

\def\0sf{\mathsf{0}}
\def\1sf{\mathsf{1}}

\def\Psf{\mathsf{P}}
\def\Qsf{\mathsf{Q}}

\def\0BS{\boldsymbol{0}}
\def\1BS{\boldsymbol{1}}

\def\0B{\mathbf{0}}
\def\1B{\mathbf{1}}

\def\0H{\hat{0}}
\def\1H{\hat{1}}

\def\sH{\hat{s}}

\def\+TT{\texttt{+}}
\def\-{\texttt{-}}
\def\+KB{|+\> \<+|}
\def\-KB{|-\> \<-|}

\def\q0{|0\>}

\def\piH{\hat{\pi}}

\def\0U{\underline{0}}
\def\1U{\underline{1}}

\def\cU{\underline{c}}

\def\fU{\underline{f}}

\def\sU{\underline{s}}

\def\xU{\underline{x}}

\def\piT{\tilde{\pi}}

\def\iO{\overline{i}}
\def\jO{\overline{j}}

\def\JO{\overline{J}}

\def\iotaU{\underline{\iota}}

\def\tauU{\underline{\tau}}

\def\piU{\underline{\pi}}

\def\rhoU{\underline{\rho}}

\def\varepsilonU{\underline{\varepsilon}}

\def\sigmaU{\underline{\sigma}}

\def\JU{\underline{J}}

\def\tOR{\mbox{ or }}

\RedeclareMathOperator{\Im}{Im}

\def\s{\underline{s}}

\def\thepage{\arabic{page}}

\newcommand{\I}{\mathbf{1}}

\def\Klove{Kl\o ve }

\theoremstyle{definition}
\newtheorem{teiri}{Theorem}

\newtheorem{lem}{Lemma}
\newtheorem{example}{Example}

\newtheorem{df}{Definition}
\newcommand{\dmin}{d_{\mathrm{min}}}
\newcommand{\dinf}{d_{\infty}}

\ifCLASSINFOpdf
\else
\fi

\begin{document}
%
\title{Recursively Extended Permutation Codes under Chebyshev Distance}
%
%
%

\author{
\IEEEauthorblockN{Tomoya Hirobe and Kenta Kasai}
\IEEEauthorblockA{Department of Information and Communications Engineering, School of Engineering\\
Institute of Science Tokyo, Tokyo 152-8550, Japan}
}

\maketitle

\begin{abstract}
We study recursively extended permutation (REP) codes under the Chebyshev distance. An REP code is built by repeatedly inserting an allowed symbol in the first coordinate and relabeling the remaining symbols. The central question is how large such a code can be for a prescribed length and minimum distance. A condition imposed separately at every extension step is sufficient to preserve distance, but it is not necessary because later extensions can increase distances. To obtain an upper bound despite this difficulty, we count the extension steps that preserve code size but are needed to remove the remaining distance shortfall. Tracking pairwise-disjoint intervals associated with codeword pairs gives a lower bound on the number of these steps. For $n>d\ge1$, this argument proves that the maximum size of a length-$n$ REP code with minimum distance at least $d$ is $\prod_{j=0}^{n-1}(\lfloor j/d\rfloor+1)$. This value equals the size of the corresponding direct product group permutation code. The recursive representation also yields a coordinate-order sequential encoder with complexity $O(n\log n)$. When the allowed insertion symbols at each step differ pairwise by at least $d$, it further yields a bounded-distance decoder with complexity $O(n\log^2 n)$.
\end{abstract}


%
\section{Introduction}
A permutation code is a set of permutations of a fixed length $n$. Introduced in the 1960s \cite{1445610}, permutation codes were later applied by Vinck et al.\ to power-line communication using $m$-ary frequency-shift keying (FSK) \cite{Vinck2000,866429}. Subsequent work studied their construction and decoding \cite{BLAKE19791,1302307,Swart2007}. In such systems, frequencies assigned to time slots represent permutation symbols, and time-frequency diversity mitigates background noise, impulse noise, and persistent frequency interference. The Chebyshev distance, induced by the $\ell_\infty$ norm, also models limited-magnitude errors in multilevel flash memories. Previous work has established Gilbert--Varshamov- and ball-packing-type bounds \cite{5466536,7342968,7908949}, efficient encoding and decoding algorithms \cite{5466546,5466536}, and systematic constructions \cite{8648459,6937135} for this metric.

Two known constructions provide the setting for our question. First, \Klove et al.\ \cite[Sec.~III.A]{5466546} and Tamo and Schwartz \cite[Construction~1]{5466536} independently constructed codes by permuting symbols separately within their residue classes modulo $d$. The resulting direct product group permutation (DPGP) code is explicit and has algebraic encoders and decoders \cite{5466536,5466546}. Its asymptotic rate is analyzed in \cite[Sec.~IV.C]{5466536}. DPGP codes are not generally maximum among all permutation codes: Bereg et al.\ proved that the maximum size at distance $2$ is $n!/2^{\lfloor n/2\rfloor}$ \cite[Theorem~7]{bereg2023improvedboundspermutationarrays}, so the length-$6$, distance-$2$ maximum is $90$, whereas the DPGP code has size $36$. Tamo and Schwartz give an extension of the DPGP construction in \cite[Construction~2]{5466536}. Han et al.\ use its right cosets to obtain a systematic construction \cite{8648459}, distinct from the systematic construction in \cite{7435295}.

Second, \Klove et al.\ introduced a code extension operation \cite[Sec.~III.C]{5466546}. It inserts an allowed symbol in the first coordinate, relabels the previous symbols, and thereby turns a permutation of length $j$ into one of length $j+1$. Repeating this operation produces a recursively extended permutation (REP) code. If $q$ symbols are allowed at one step, that step multiplies the number of codewords by $q$. This transparent size growth suggests the extremal question studied here: among length-$n$ REP codes with minimum Chebyshev distance at least $d$, what is the largest possible number of codewords?

A direct answer cannot be obtained by imposing the distance requirement separately at each step. Requiring the allowed symbols at every step to differ pairwise by at least $d$ preserves minimum distance, but this condition is not necessary. A later extension with only one allowed symbol can increase distances that were previously below $d$. Such a step improves distance but does not increase the number of codewords. The upper-bound problem therefore requires an accounting of how many size-preserving steps are forced by earlier choices that increased the code size.

We carry out this accounting in two stages. For an intermediate code $C$, we define $c_1(C;d)$ as the minimum number of future singleton-head extensions needed to reach minimum distance $d$. For a head set $S$, we define $\delta_d(S)$ as the total shortfall of its consecutive spacings from $d$. By following pairwise-disjoint maximum intervals of carefully chosen codeword pairs, we prove that extending by $S$ creates a future requirement of at least $\delta_d(S)$ singleton-head steps. Combining this lower bound with an upper bound on the number of heads that can be placed with a given total spacing shortfall yields
\begin{align*}
\max_{\substack{\Cc^{(n)}\text{ REP}\\ \dmin(\Cc^{(n)})\ge d}}
|\Cc^{(n)}|
=\prod_{j=0}^{n-1}\left(\left\lfloor\frac{j}{d}\right\rfloor+1\right),
\qquad n>d\ge1.
\end{align*}
Equally spaced head sets attain the bound. The product is also the cardinality of the corresponding DPGP code, although the two codes need not have the same pairwise-distance structure.

The recursive description also supports efficient algorithms. \Klove et al.\ gave encoding and decoding algorithms for the case of two allowed heads per step \cite[Sec.~III.D]{5466546}. Kawasumi and Kasai subsequently used the encoder's tree-structured factor graph for maximum a posteriori (MAP) decoding \cite{19950201} and studied concatenation with low-density parity-check (LDPC) codes \cite{KawasumiKasai2023}. For arbitrary head sets, we show that reading the head sequence in reverse gives the ranks used in standard permutation unranking. This observation yields a coordinate-order encoder with complexity $O(n\log n)$. When every head set is pairwise $d$-separated, the same rank representation gives a bounded-distance decoder with complexity $O(n\log^2 n)$.

The argument proceeds from the comparison value to the obstruction and then to its resolution. Section~II fixes the notation and derives the DPGP cardinality that will reappear in the main theorem. Section~III analyzes one extension step: first its effect on size and distance, and then the interval structure needed to follow distance growth. Section~IV converts those local facts into a lower bound on the number of forced size-preserving steps, proves the maximum-size theorem, and gives an attaining construction. Section~V uses the same recursive representation for encoding and decoding. Section~VI summarizes the result and its scope.
\section{Notation and Preliminaries}
The maximum-size theorem compares an REP code with a DPGP code having the same length and target distance. We first establish a common notation for permutation codes and then derive the DPGP cardinality in the product form that will appear again as the REP upper bound.

For a positive integer $n$, let $[n]:=\{0,1,\ldots,n-1\}$.

Let \(\Sc_n\) denote the symmetric group on \([n]\), that is, the set of bijections \(f:[n]\to[n]\). More generally, for a finite set $A$, let $\Sc_A$ denote the symmetric group on $A$. We represent \(f\in\Sc_n\) by the array \(\fU=[f(0),\ldots,f(n-1)]\). Arrays are denoted by underlined symbols; for example, \(\xU=[x_0,\ldots,x_{n-1}]\), where \(x_j\) is the component at coordinate \(j\). We also set \(\Sc_0:=\{\varepsilonU\}\), where \(\varepsilonU\) is the empty permutation.

A nonempty subset \(C\subset\Sc_n\) is a \emph{permutation code} of length \(n\), and its elements are \emph{codewords}. For \(\cU,\cU'\in C\), their Chebyshev distance is
\[
d_\infty(\cU, \cU') = \max_{j \in [n]} |c_j - c_j'|.
\]
Thus two permutations are close when their symbols differ only slightly at every coordinate; their distance is the largest coordinatewise difference.
The minimum distance of \(C\) is
\[
d_\infty(C) := \min_{\cU, \cU' \in C : \cU \neq \cU'} d_\infty(\cU, \cU').
\]
For a singleton code, we set \(d_\infty(C)=\infty\). An \((n,M,d)\) code is a length-\(n\) code of size \(M\) and minimum distance at least \(d\).

\subsection{Direct Product Group Permutation Codes}\label{212515_27Aug24}
The DPGP code supplies the numerical benchmark for the REP extremal problem. Its symbols can move only within residue classes modulo $d$, which guarantees minimum distance at least $d$ and makes its cardinality explicit.

We review the construction independently introduced by \Klove et al.\ \cite[Sec.~III.A]{5466546} and Tamo and Schwartz \cite[Construction~1]{5466536}. Fix integers $n,d\ge1$. For \(i\in[d]\), let
\( A_i = (d\mathbb{Z} + i) \cap [n] = \{j \in [n] \mid j \equiv i \ (\bmod\ d)\}. \)
The corresponding DPGP code is
\(G:=\{\piU\in\Sc_n:\pi_i\equiv i\pmod d\text{ for every }i\in[n]\}\).
Equivalently, it is the direct product of the symmetric groups on the residue classes:
\( G = \Sc_{A_0} \times \Sc_{A_1} \times \cdots \times \Sc_{A_{d-1}}. \)
If $\piU,\sigmaU\in G$ are distinct, then at some coordinate $j$ the nonzero difference $\pi_j-\sigma_j$ is divisible by $d$. Hence $\dmin(G)\ge d$. When $n>d$, swapping $0$ and $d$ in the identity permutation gives a codeword at distance $d$ from the identity, so $\dmin(G)=d$.
{\begin{example}
For $n=6$ and $d=2$, the residue classes are $A_0=\{0,2,4\}$ and $A_1=\{1,3,5\}$, so $G=\Sc_{A_0}\times\Sc_{A_1}$. A codeword is obtained by independently permuting each class and interleaving the results in their fixed coordinate positions. For example, $[024]\in\Sc_{A_0}$ and $[135]\in\Sc_{A_1}$ produce $[012345]\in G$, whereas $[420]\in\Sc_{A_0}$ and $[531]\in\Sc_{A_1}$ produce $[452301]\in G$. Hence
$|G| = |\Sc_{A_0}| \cdot |\Sc_{A_1}| = 3! \cdot 3! = 36$.
\end{example}
}
The size of \(A_i\) is \(\lfloor n/d\rfloor\) when \(i\ge n\bmod d\) and \(\lceil n/d\rceil\) otherwise. Consequently,
\[
|G|=\left(\left\lceil \frac{n}{d} \right\rceil!\right)^{n \bmod\ d}
\left(\left\lfloor \frac{n}{d} \right\rfloor!\right)^{d - (n \bmod\ d)}.
\]
When \(d\) divides \(n\), this reduces to \(|G|=((n/d)!)^d\).

This derivation follows \cite[Construction~1]{5466536}; see also \cite[Sec.~III.A]{5466546}.

The same cardinality has the product representation
\begin{align}
|G|=\prod_{j=0}^{n-1}\left(\left\lfloor\frac{j}{d}\right\rfloor+1\right). \label{dpgp_product}
\end{align}
This form records one factor for each length increase. It is therefore the form that can be compared directly with a recursively constructed code.
Indeed, write $n=qd+r$ with $0\le r<d$. Grouping the first $qd$ factors into $q$ blocks of length $d$ gives
\begin{align*}
\prod_{j=0}^{n-1}\left(\left\lfloor\frac{j}{d}\right\rfloor+1\right)
&=\prod_{p=0}^{q-1}(p+1)^d\,(q+1)^r \\
&=(q!)^d(q+1)^r \\
&=\left(\left\lceil\frac{n}{d}\right\rceil!\right)^r
  \left(\left\lfloor\frac{n}{d}\right\rfloor!\right)^{d-r}.
\end{align*}
\section{One-Step Code Extension}
An REP code is a sequence of local extension choices, so its global size and distance can be understood only after one extension step is analyzed exactly. Three questions matter: how many codewords the step creates, which distances it preserves, and when it can increase a distance that is still below the target. The first two questions give immediate construction rules; the third leads to the interval analysis used in the upper bound.
\subsection{Extension Operation and Head Sets}
The extension operation introduced by \Klove et al.\ \cite[Sec.~III.C]{5466546} inserts a new first symbol while preserving the relative order of all previous symbols. The allowed first symbols determine both the number of new codewords and part of their separation.

Let $\piU=[\pi_0,\dots,\pi_{n-1}]\in\Sc_n$, where $n\ge1$. Let $s\in[n+1]$ be a symbol to be inserted in the first coordinate; following \cite{5466546}, we call $s$ a \emph{head}. The \emph{extension} of $\piU$ by $s$ is the permutation
\begin{align}
 \piU^s := \bigl[s, \pi_0^s, \pi_1^s, \dots, \pi_{n-1}^s\bigr], \label{001517_13Apr23}
\end{align}
where $x^s:=\phi_s(x):=x+\I[x\ge s]$ and $\I[P]$ is the indicator of proposition $P$. The weak inequality is necessary: the strict inequality printed in \cite[Sec.~III.C]{5466546} does not always produce a permutation.
In words, the new symbol $s$ is placed first, and every old symbol at least $s$ is increased by one to leave a vacant value for it.

For a code $\Cc\subset\Sc_n$, let $S\subset[n+1]$ be the set of allowed heads, which we call the \emph{head set}. Define the extended code
\[
\Cc^S := \{\piU^s \in \Sc_{n+1} \mid s \in S, \piU \in \Cc\}.
\]
We also write $\phi(\piU;s):=\piU^s$ and $\phi(\Cc;S):=\Cc^S$.
All head sets are assumed to be nonempty.

For the empty permutation, define
\[
\varepsilonU^0 = [0].
\]

For a head set $S\subset[n+1]$, define
\[
\dmin(S) \defeq \min \bigl\{ |s - s'| : s, s' \in S, s \neq s' \bigr\}.
\]
For $|S|=1$, set $\dmin(S)=\infty$.
The quantity $\dmin(S)$ is the smallest separation between two allowed first symbols. It will give an immediate lower bound on the distance between codewords created from different heads.

\begin{example}
For $\piU=[0123]$ and $s=2$, the extension is $\piU^s=[0123]^2=[20134]$.
For $\Cc=\{[0123]\}$ and $S=\{0,2,4\}$, we have $\Cc^S=\{[01234],\allowbreak [20134],[40123]\}. $
\end{example}

\subsection{Order, Size, and Elementary Distance Properties}
The extension relabeling is order preserving. This elementary fact keeps interval relations intact and will later allow several distance constraints to be followed simultaneously.

\begin{lem}\label{144109_3Jun24}
For $\pi,\sigma\in [n]$ and $s\in [n+1]$, 
\begin{enumerate*}[label=\roman*), series = tobecont, itemjoin = \quad]
\item \label{145251_3Jun24}
$\pi < \sigma$ implies $\pi^s < \sigma^s$.
\item \label{145256_3Jun24}
$\pi \le \sigma$ implies $\pi^s \le \sigma^s$.
\end{enumerate*}
\end{lem}
\begin{IEEEproof}\label{144130_3Jun24}
For part~\ref{145251_3Jun24}, the three possible cases are
$\pi^s=\pi+1<\sigma+1=\sigma^s$ if $s\le\pi<\sigma$,
$\pi^s=\pi<\sigma=\sigma^s$ if $\pi<\sigma<s$, and
$\pi^s=\pi<\sigma+1=\sigma^s$ if $\pi<s\le\sigma$.
Part~\ref{145256_3Jun24} follows from part~\ref{145251_3Jun24} and the equality $\pi^s=\sigma^s$ when $\pi=\sigma$.
\end{IEEEproof}

Before considering a code, we record the maximum possible number of pairwise $d$-separated heads. This is the largest multiplication factor available to one extension when the target separation is imposed at that step.
\begin{teiri}\label{230931_30Oct23}
For any integer $d\ge1$ and any $S\subset[n+1]$ with $\dmin(S)\ge d$,
$d(|S|-1)\le n$ and hence $|S|\le\lfloor n/d\rfloor+1$.
Conversely, the set $S = \{0,d,2d,\ldots,\lfloor n/d\rfloor d\}\subset[n+1]$ has $|S|=\lfloor n/d\rfloor+1$ and $\dmin(S)\ge d$; its minimum spacing equals $d$ when $|S|\ge2$. 
\end{teiri}
\begin{IEEEproof}\label{001824_29Oct23}
Write $S=\{s_1<\cdots<s_{|S|}\}$. The sets $\{s_1\}$ and $(s_i,s_{i+1}]$, $1\le i<|S|$, are pairwise disjoint and their union is contained in $[n+1]$. Therefore,
\[1+d(|S|-1)\le1+\sum_{i=1}^{|S|-1}|s_i-s_{i+1}|\le n+1.\]
The stated construction attains the cardinality bound and has adjacent spacing $d$ whenever it contains at least two elements.
\end{IEEEproof}

\begin{example}\label{141422_10Jan24}
For $n=5$, $S=\{0,3,5\}$, we have $\dmin(S)=2$, $|S|=3$, and $\lfloor n/\dmin(S)\rfloor+1=\lfloor5/2\rfloor+1=3$. 
For $n=6$, $S=\{0,3,6\}$, we have $\dmin(S)=3$, $|S|=3$, and $\lfloor n/\dmin(S)\rfloor+1=\lfloor6/3\rfloor+1=3$. 
\end{example}

If the underlying permutation is fixed and only the head changes, the first coordinate determines the entire Chebyshev distance.
\begin{lem}\label{010525_27Nov23}
For a permutation $\piU\in\Sc_n$ and $s,t\in[n+1]$,
\begin{align}
\dinf(\piU^s,\piU^t)=|s-t|.
\end{align}
\end{lem}

\begin{IEEEproof}\label{012433_27Nov23} 
If $s=t$, both sides are zero. If $s\ne t$, then $|s-t|\ge1$ and $|\pi_j^s-\pi_j^t|\le1$ for every $j\in[n]$. The head coordinates differ by $|s-t|$, and hence
$\dinf(\piU^s,\piU^t)=\max\{|s-t|,|\pi_j^s-\pi_j^t|:j\in[n]\}=|s-t|$.
\end{IEEEproof}

To describe when relabeling increases a symbol difference, we use the integer values lying strictly above the smaller endpoint and at most the larger endpoint.
\begin{df}[Interval]\label{171541_8Apr25}
For nonnegative integers $x$ and $y$, define the \emph{interval between $x$ and $y$} by
${\Ic}(x,y):=(\min\{x,y\},\max\{x,y\}]$ when $x\ne y$, and ${\Ic}(x,x):=\emptyset$.
\end{df}

An extension increases a symbol difference by one exactly when its head lies between the two symbols. This identity is the local distance rule used throughout the later arguments.
\begin{lem}\label{lem:phi_kakudai}
For $n>0$, $s\in[n+1]$, and $\pi,\sigma\in[n]$,
\begin{align}
\left| \pi^s - \sigma^s \right| = \left| \pi - \sigma \right| + \I[s \in \Ic(\pi,\sigma)]. \label{142940_10Apr23}
\end{align}
\end{lem}
\begin{IEEEproof}
\begin{align}
  |\pi^s-\sigma^s|
&= \left|\phi_s\left(\pi\right)-\phi_s\left(\sigma\right)\right| 
\\&=\left|\left(\pi-\sigma\right)+\left(\I[\pi \geq s]-\I[\sigma \geq s]\right)\right| 
\\&= |\pi-\sigma|+\I[\sigma< s\le \pi \tOR \pi< s\le \sigma]
\\&= |\pi-\sigma|+\I[s\in \Ic(\pi,\sigma)]. 
\end{align}
\end{IEEEproof}

Applying the same head to two codewords therefore never decreases their distance and can increase it by at most one.
\begin{lem}\label{224431_12Nov23}
For a length-$n$ code $\Cc$, distinct $\piU,\sigmaU\in\Cc$, and $s\in[n+1]$,
$\dinf(\piU, \sigmaU) \le d_\infty\bigl(\piU^s, \sigmaU^s\bigr) \le \dinf(\piU, \sigmaU) + 1. $
\end{lem}
\begin{IEEEproof}
The head coordinates of $\piU^s$ and $\sigmaU^s$ are equal. For each remaining coordinate, Lemma~\ref{lem:phi_kakudai} gives
$|\pi_j-\sigma_j|\le|\pi_j^s-\sigma_j^s|\le|\pi_j-\sigma_j|+1$.
Taking the maximum over $j\in[n]$ proves the claim.
\end{IEEEproof}

Applying different heads gives another source of separation: the new first coordinate alone separates the extended codewords by the distance between the heads.
\begin{lem}[Distance lower bound from head symbols]\label{143827_10May23}
For a code $\Cc\subset\Sc_n$, arbitrary $\piU,\sigmaU\in\Cc$, and $s,t\in[n+1]$,
$d_\infty\bigl(\piU^s, \sigmaU^t\bigr) \ge |s - t|$.
Equality holds when $\piU = \sigmaU$.
\end{lem}
\begin{IEEEproof}
\begin{align*}
d_\infty\left(\piU^s, \sigmaU^t\right)
&= \max_{j \in [n+1]} \left| (\piU^s)_j - (\sigmaU^t)_j \right| \\
&\geq \left| (\piU^s)_0 - (\sigmaU^t)_0 \right| \\
&= \left| s - t \right|.
\end{align*}
Equality for $\piU=\sigmaU$ follows from Lemma~\ref{010525_27Nov23}.
\end{IEEEproof}

Together, the preceding bounds show that neither the original codeword nor the chosen head is lost in the extension.
\begin{lem}\label{152042_1Oct23}
The map $\phi:\Sc_n\times[n+1]\to\Sc_{n+1}$ is injective.
\end{lem}
\begin{IEEEproof}
Suppose $(\piU,s)\ne(\sigmaU,t)$. If $s\ne t$, Lemma~\ref{143827_10May23} gives $\piU^s\ne\sigmaU^t$. If $s=t$ and $\piU\ne\sigmaU$, choose $i\in[n]$ with $\pi_i\ne\sigma_i$. Lemma~\ref{lem:phi_kakudai} gives $|\phi_s(\pi_i)-\phi_s(\sigma_i)|\ge|\pi_i-\sigma_i|>0$, so again $\piU^s\ne\sigmaU^t$.
\end{IEEEproof}

Consequently, every original codeword produces exactly one distinct extended codeword for each allowed head.
\begin{teiri}\label{205810_6Jul23}
For a code $\Cc\subset\Sc_n$ and a head set $S\subset[n+1]$,
$|\Cc^S|=|\Cc||S|$.
\end{teiri}
\begin{IEEEproof}
The claim follows because the map $\phi$ is injective by Lemma~\ref{152042_1Oct23}.
\end{IEEEproof}

\subsection{Lower Bounds on Minimum Distance under Extension}
The two sources of separation identified above combine into a simple construction rule. Codewords obtained from the same head remain at least as far apart as the original codewords, while codewords obtained from different heads are separated by the head spacing.

\begin{teiri}\label{153755_6Jan24}
For any code $\Cc\subset\Sc_n$ and any head set $S\subset[n+1]$,
\begin{align}
\dmin\left(\Cc^S\right) \geq \min \left(\dmin(S), \dmin(\Cc)\right).
\end{align}
\end{teiri}
\begin{IEEEproof}\label{153804_6Jan24}
Suppose first that $S=\{s\}$, so $\dmin(S)=\infty$. Every distinct pair in $\Cc^S$ has the form $(\piU^s,\sigmaU^s)$ for distinct $\piU,\sigmaU\in\Cc$.
Lemma~\ref{224431_12Nov23} gives $\dinf(\piU^s,\sigmaU^s)\ge\dinf(\piU,\sigmaU)\ge\dmin(\Cc)$. Hence $\dmin(\Cc^S)\ge\dmin(\Cc)=\min\{\dmin(S),\dmin(\Cc)\}$.

Now suppose $|S|\ge2$ and consider distinct codewords $\piU^s,\sigmaU^t\in\Cc^S$.
\begin{itemize}
\item
If $s\ne t$, Lemma~\ref{143827_10May23} gives $\dinf(\piU^s,\sigmaU^t)\ge|s-t|\ge\dmin(S)$.
\item
If $s=t$, then $\piU\ne\sigmaU$, and Lemma~\ref{224431_12Nov23} gives $\dinf(\piU^s,\sigmaU^s)\ge\dinf(\piU,\sigmaU)\ge\dmin(\Cc)$.
\end{itemize}
The desired lower bound follows in both cases.
\end{IEEEproof}

Equivalently, once both the original code and the head set have minimum distance at least $d$, one extension preserves that target distance. Repeating this statement gives the standard sufficient condition for constructing REP codes.
\begin{teiri}[{\cite[Theorem 4]{5466546}}]\label{163227_14May23}
For an integer $d\ge1$, a code $\Cc\subset\Sc_n$, and a head set $S\subset[n+1]$, if $\dmin(S)\ge d$ and $\dmin(\Cc)\ge d$, then $\dmin(\Cc^S)\ge d$.
\end{teiri}
\begin{IEEEproof}\label{161444_6Jan24}
The assumptions give
\[
\min\{\dmin(S),\dmin(\Cc)\}\ge d.
\]
The result follows from Theorem~\ref{153755_6Jan24}.
\end{IEEEproof}

\subsection{Upper Bounds on Minimum Distance under Extension}
The preceding lower bound explains how to guarantee distance, but an extremal proof also needs to know how much one extension can improve it. Two complementary bounds limit the new minimum distance: one comes from the closest pair of heads, and the other from the old minimum distance plus one.

\begin{teiri}[Upper bound on $\dmin(\Cc^S)$]\label{003746_1Oct23}
Let $\Cc\subset\Sc_n$ be a code and $S\subset[n+1]$ a head set. Then
\begin{align}
\dmin(\Cc^S) \le \dmin(S).
\end{align}
\end{teiri}
\begin{IEEEproof}
For $|S|=1$, the bound is immediate because $\dmin(S)=\infty$. Suppose $|S|\ge2$.
Select $s, t \in S$ such that $|s-t| = \dmin(S)$.
For any $\piU\in\Cc$, Lemma~\ref{010525_27Nov23} gives $d_\infty(\piU^s,\piU^t)=|s-t|=\dmin(S)$.
Thus $\Cc^S$ contains a pair at distance $\dmin(S)$.
\end{IEEEproof}

Even if the heads are far apart, a pair of old codewords extended by the same head can gain at most one unit of distance.
\begin{teiri}\label{161110_1Oct23}
Let $\Cc\subset\Sc_n$ be a code and $S\subset[n+1]$ a head set. Then $d_{\infty}(\Cc^S)\le\dmin(\Cc)+1$.
\end{teiri}
\begin{IEEEproof}
If $|\Cc|=1$, then $\dmin(\Cc)=\infty$ and the claim is immediate. Suppose $|\Cc|\ge2$.
Select distinct $\piU, \sigmaU \in \Cc$ such that $d_\infty(\piU, \sigmaU) = \dmin(\Cc)$.
For any $s\in S$, Lemma~\ref{152042_1Oct23} shows that $\piU^s$ and $\sigmaU^s$ are distinct, while Lemma~\ref{224431_12Nov23} gives $\dinf(\piU^s,\sigmaU^s)\le\dmin(\Cc)+1$.
This pair gives the claimed upper bound.
\end{IEEEproof}

By Theorem~\ref{205810_6Jul23}, an extension is \emph{size-preserving} when $|S|=1$ and \emph{size-increasing} when $|S|>1$. It is \emph{distance-increasing} when $\dmin(\Cc^S)>\dmin(\Cc)$.

\begin{example}\label{134606_2Oct23}
Let $\Cc=\{[0123],[3012]\}$ and $S=\{1\}$. Then $\Cc^S=\{[10234],[14023]\}$, with
\[
\dmin(S) = \infty, \quad \dmin(\Cc) = 3, \quad \text{and} \quad \dmin(\Cc^S) = 4.
\]
Thus, the extension is size-preserving and distance-increasing.

For $\Cc=\{[0123],[1032]\}$ and $S=\{1,3\}$,
\[
\begin{aligned}
\Cc^S &= \{[10234], [30124], [12043], [31042]\},\\
\dmin(S) &= 2, \quad \dmin(\Cc) = 1, \quad \dmin(\Cc^S) = 2.
\end{aligned}
\]
Thus, the extension is both size-increasing and distance-increasing.
\end{example}

\subsection{Codeword Pairs, Interval Sets, and Maximum Intervals}
The upper bounds reveal the central difficulty: a singleton-head extension can repair a distance shortfall without increasing code size. To count how many such repairs are necessary, we need to know which head values can increase the distance of a given codeword pair. Intervals provide exactly this information and, unlike the distance value alone, can be followed through later extensions.

The length of $I={\Ic}(x,y)$ is $|I|:=|x-y|$. For $x,y\in[n]$ and $s\in[n+1]$, define its extension by
$I^s:={\Ic}(x^s,y^s)$.  

An interval gains one element precisely when the inserted head lies inside it.
\begin{lem}\label{054652_14Jul24} 
For an interval $I={\Ic}(x,y)$ with $x,y\in[n]$ and a head $s\in[n+1]$,
$|I^s|=|I|+\I[s\in I]$.
\end{lem}
\begin{IEEEproof}
By Lemma~\ref{lem:phi_kakudai},
$|I^s|=|{\Ic}(x^s,y^s)|=|x^s-y^s|=|x-y|+\I[s\in {\Ic}(x,y)]=|I|+\I[s\in I]$. 
\end{IEEEproof}
{\begin{example}
 Let $I = (1,4]$, so $|I| = 3$. For $s = 2 \in I$, we have
 $I^s = (\phi_2(1), \phi_2(4)] = (1,5] = \{2,3,4,5\}, \quad |I^s| = 4 = |I| + 1$.
\end{example}
}
Distinct distance shortfalls can be tracked separately because disjoint intervals remain disjoint under the combinations of extensions used later.
\begin{lem}\label{153103_22May24}
Let $I={\Ic}(x_1,y_1)$ and $J={\Ic}(x_2,y_2)$ be nonempty intervals with endpoints in $[n]$, and let $s,t\in[n+1]$. If $I$ and $J$ are disjoint, then:
\begin{enumerate}[label=\roman*)]
    \item \label{045307_11Apr25} For any head $s$, the intervals $I^s$ and $J^s$ are disjoint.
    \item \label{045314_11Apr25} If $s \in I$ and $t \in J$, then $I^t$ and $J^s$ are disjoint.
    \item \label{045321_11Apr25} If $s \in I$, then $I^t$ and $J^s$ are disjoint.
\end{enumerate}
\end{lem}
\begin{IEEEproof}
For part~\ref{045307_11Apr25}, assume without loss of generality that $I=\Ic(\pi_1,\sigma_1)$ and $J=\Ic(\pi_2,\sigma_2)$ with $\pi_1<\sigma_1\le\pi_2<\sigma_2$.
Lemma~\ref{144109_3Jun24} gives $\pi_1^s<\sigma_1^s\le\pi_2^s<\sigma_2^s$, so $I^s$ and $J^s$ are disjoint.

For part~\ref{045314_11Apr25}, assume without loss of generality that $\pi_1<s\le\sigma_1\le\pi_2<t\le\sigma_2$.  
Then $I^t = \Ic(\pi_1^t, \sigma_1^t) = \Ic(\pi_1, \sigma_1)$ and $J^s = \Ic(\pi_2^s, \sigma_2^s) = \Ic(\pi_2 + 1, \sigma_2 + 1)$.  
These intervals are disjoint.

For part~\ref{045321_11Apr25}, first suppose $I$ lies to the left of $J$, so $\pi_1<s\le\sigma_1\le\pi_2<\sigma_2$. If $I^t$ and $J^s$ intersected, then $\pi_2^s<\sigma_1^t$. However, $\pi_2^s=\pi_2+1$ and $\sigma_1^t\le\sigma_1+1$, which would imply $\pi_2<\sigma_1$, contradicting $\sigma_1\le\pi_2$. If $J$ lies to the left of $I$, write $J=(\pi_2,\sigma_2]$ and $I=(\pi_1,\sigma_1]$ with $\sigma_2\le\pi_1<s$. Then $J^s=J$, whereas the lower endpoint of $I^t$ is $\pi_1^t\ge\pi_1\ge\sigma_2$, so the intervals are again disjoint.
\end{IEEEproof}
{\begin{example}
 Let $I=(1,3]$, $J=(4,5]$, and $s=2\in I$.
 \begin{enumerate}[label=\roman*)]
  \item $I^2=(1,4]=\{2,3,4\}$ and $J^2=(5,6]=\{6\}$ are disjoint.
  \item For $t=5\in J$, $I^t=(1,3]$ and $J^s=(5,6]$ are disjoint.
  \item For $t=0$, $I^t=(2,4]$ and $J^s=(5,6]$ are disjoint.
 \end{enumerate}
\end{example}
}

Containment is also preserved, so an interval that controls all coordinate differences before extension continues to control them afterward.
\begin{lem}\label{195413_3Jun24}
If nonempty intervals $I$ and $J$ satisfy $I\subset J$, then $I^s\subset J^s$.
\end{lem}

\begin{IEEEproof}
Without loss of generality, write $I = \Ic(\pi_1, \sigma_1)$ and $J = \Ic(\pi_2, \sigma_2)$ with $\pi_2 \le \pi_1 < \sigma_1 \le \sigma_2$.
From Lemma~\ref{144109_3Jun24}, $\pi_2^s \le \pi_1^s < \sigma_1^s \le \sigma_2^s$, so $I^s \subset J^s$.
\end{IEEEproof}
For distinct permutations $\piU,\sigmaU\in\Sc_n$, define the following objects.
\begin{enumerate}
\item
The \emph{interval set} between $\piU$ and $\sigmaU$ is
$\Ic(\piU,\sigmaU):=\{\Ic(\pi_j,\sigma_j):\pi_j\ne\sigma_j,\ j\in[n]\}$.
\item
An interval in $\Ic(\piU,\sigmaU)$ that contains every other interval in this set is the \emph{maximum interval} of the pair. If it exists, it is unique.
\end{enumerate}
The interval set records the value gap at every coordinate where the permutations differ. A maximum interval, when it exists, contains all those gaps; its length is therefore the Chebyshev distance of the pair.
By definition,
\[
\Ic(\piU^s, \sigmaU^s)
=\{\Ic(\pi_j^s,\sigma_j^s):\pi_j\ne\sigma_j,\ j\in[n]\}.
\]
and
\[
\dinf(\piU, \sigmaU) = \max_{J \in \Ic(\piU, \sigmaU)} |J|.
\]

For a pair $\Psf:=(\piU,\sigmaU)\in\Sc_n^2$ and $s\in[n+1]$, write $\Psf^s:=(\piU^s,\sigmaU^s)$.
Maximum intervals identify both the current distance and the exact heads that increase it, as stated below.
\begin{lem}\label{060127_4Mar24}
Suppose $\Psf:=(\piU,\sigmaU)\in\Sc_n^2$ has maximum interval $I$. Then:
\begin{enumerate}[label=\roman*)]
\item\label{142848_5Mar24}
The permutation pair $\Psf^s$ has maximum interval $I^s$. 
\item\label{051045_10Apr25}
$\dinf(\piU, \sigmaU) = |I|${.}
\item\label{153314_30Dec23}
$\dinf(\piU^s, \sigmaU^s) = \dinf(\piU, \sigmaU) + \I[s \in I]${.} 
\item\label{153314b_30Dec23}
$|I^s| = |I| + \I[s \in I]{.}$
 \end{enumerate}
\end{lem}
\begin{IEEEproof}\label{162409_30Dec23}
For part~\ref{142848_5Mar24}, every $J\in\Ic(\piU,\sigmaU)$ satisfies $J\subset I$. Lemma~\ref{195413_3Jun24} therefore gives $J^s\subset I^s$, so $I^s$ is the maximum interval of $\Psf^s$. Part~\ref{051045_10Apr25} follows from the definition of Chebyshev distance. Lemma~\ref{054652_14Jul24} gives part~\ref{153314b_30Dec23}. If $s\in I$, the length of $I$ increases by one; if $s\notin I$, then $s$ lies in no interval contained in $I$, so no interval length increases. Together with part~\ref{142848_5Mar24}, this proves part~\ref{153314_30Dec23}.
\end{IEEEproof}
\begin{example}
Let $n=5$, $\piU=[01234]$, and $\sigmaU=[01432]$. The maximum difference is $|\pi_4-\sigma_4|=|4-2|=2$, so the maximum interval is $I=(2,4]$. For $s=3\in I$,
$\piU^3 = [301245], \quad \sigmaU^3 = [301542]$,
and $I^3=(2,5]$. Hence
$d_\infty(\piU^3, \sigmaU^3) = 3 = d_\infty(\piU, \sigmaU) + 1$ and $|I^3| = 3 = |I| + 1$.
\end{example}

Finally, two codewords created from the same original permutation by different heads have a maximum interval determined solely by those heads. This supplies the initial disjoint intervals used in the counting argument of Section~IV.
\begin{lem}\label{000804_8Dec23} 
Let $\piU\in\Sc_n$ and $s,t\in[n+1]$ with $s<t$. Then:
\begin{enumerate}[label=\roman*)]
 \item \label{133646_30Dec23}
 $ \Ic(\piU^s,\piU^t)=\{{\Ic}(s,t),{\Ic}(s,s+1),\ldots,{\Ic}(t-1,t)\}$.
 \item 
The two extended permutations differ in exactly $|s-t|+1$ coordinates.
 \item 
 The codeword pair $(\piU^s,\piU^t)$ has maximum interval ${\Ic}(s,t)$.
 \end{enumerate}
\end{lem}
\begin{IEEEproof}
Reordering the coordinates does not change the interval set or the number of differing coordinates, so assume without loss of generality that $\piU$ is the identity permutation $\iotaU=[0,1,\ldots,n-1]$. By \eqref{001517_13Apr23},
\begin{align}
\begin{aligned}
\iotaU^s={}&[\black{s},0,1,\ldots,s-1,\black{s+1},\\
            &\phantom{[}\black{s+2,\ldots,t},t+1,\ldots,n],\\
\iotaU^t={}&[\black{t},0,1,\ldots,s-1,\black{s},\\
            &\phantom{[}\black{s+1,\ldots,t-1},t+1,\ldots,n].
\end{aligned}
\end{align}
The displayed arrays give
 \begin{align*}
    \Ic(\piU^s,\piU^t)=\{{\Ic}(s,t),{\Ic}(s,s+1),\ldots,{\Ic}(t-1,t)\}.
    \end{align*}
The arrays differ in the head coordinate and in the $t-s$ coordinates occupied by $s,\ldots,t-1$, hence in exactly $|s-t|+1$ coordinates. Different coordinates can determine the same interval; for example, this occurs when $t=s+1$. Therefore, the interval set can contain fewer than $|s-t|+1$ elements. Finally, ${\Ic}(s,t)$ contains every other displayed interval and is therefore the maximum interval.
\end{IEEEproof}

{\begin{example}
Let $n=5$, $\piU=[01234]$, $s=1$, and $t=3$. Then
\[
\piU^s=[102345], \qquad \piU^t=[301245].
\]
Their interval set is $\Ic(\piU^s,\piU^t)=\{(1,3],(1,2],(2,3]\}$. The permutations differ in $3=|t-s|+1$ coordinates, and their maximum interval is $(s,t]=(1,3]$.
\end{example}
}

\section{REP Codes and Their Maximum Size}\label{164043_17Sep24}
Repeated extension makes the code size a product of local head-set sizes. The difficulty is that minimum distance is global: a step that creates close codewords may be repaired later by extensions that do not enlarge the code. We first define the REP family and exhibit this delayed repair, then measure its unavoidable cost and convert that cost into the exact maximum size.

For each $j=0,\ldots,n-1$, let $S^{(j)}$ be a nonempty subset of $[j+1]$. Starting from $\Cc^{(0)}:=\{\varepsilonU\}$, define, for $j=1,\ldots,n$,
$\Cc^{(j)} = \phi(\Cc^{(j-1)}; S^{(j-1)})$.
The resulting \emph{recursively extended permutation} (REP) code is denoted by $\Cc^{(n)}=\<S^{(0)},\ldots,S^{(n-1)}\>$. Theorem~\ref{205810_6Jul23} gives
$|\Cc^{(n)}| = \prod_{j=0}^{n-1} |S^{(j)}|$.
Thus a message chooses one head from each set, and every head sequence produces one codeword. The cardinality is large when many stages offer several choices.

\begin{example}
For integers $n,d,q$ with $q\ge2$ and $(q-1)d<n$, \cite[Sec.~III.D]{5466546} constructs an $(n,q^{n-(q-1)d},d)$ REP code as follows. Set $S^{(j)}=\{0\}$ for $0\le j<(q-1)d$, and set
$S^{(j)}=\{\lfloor j/(q-1)\rfloor x:x=0,\ldots,q-2\}\cup\{j\}$ for $(q-1)d\le j<n$. The latter sets contain $q$ elements with minimum spacing at least $d$. Therefore,
$|\Cc^{(n)}|=\prod_{j=0}^{n-1}|S^{(j)}|=q^{n-(q-1)d}$, and repeated application of Theorem~\ref{163227_14May23} gives $\dmin(\Cc^{(n)})\ge d$.
\end{example}

{\begin{example}\label{ex:example8}
Let $n=7$, $d=2$, and $q=3$. Then $(q-1)d=4<n$, and the head sets are
\begin{align*}
S^{(0)} &= \{0\}, \quad
S^{(1)} = \{0\}, \quad
S^{(2)} = \{0\}, \quad
S^{(3)} = \{0\}, \\
S^{(4)} &= \{0,2,4\}, \quad
S^{(5)} = \{0,2,5\}, \quad
S^{(6)} = \{0,3,6\}.
\end{align*}
The resulting REP code has size
\[
|\Cc^{(7)}| = 1 \cdot 1 \cdot 1 \cdot 1 \cdot 3 \cdot 3 \cdot 3 = 27 = 3^{7 - (3-1)\cdot 2}.
\]
Each $S^{(j)}$ for $j\ge4$ has minimum spacing at least $2$, so Theorem~\ref{163227_14May23} gives $\dmin(\Cc^{(7)})\ge2$.
\end{example}
}
Theorem~\ref{163227_14May23} ensures $\dmin(\Cc^{(n)})\ge d$ if $\dmin(S^{(j)})\ge d$ for every $j$. The converse fails: an intermediate head set may have minimum distance below $d$ even when the final REP code has minimum distance at least $d$. This distinction is why a stage-by-stage spacing bound cannot prove the desired upper bound.
For instance, consider $S^{(0)}=\{0\}$, $S^{(1)}=\{0,1\}$, and $S^{(2)}=\{1\}$, where $\dmin(S^{(1)})=1$. Then
$\Cc^{(0)}=\{\varepsilonU\}$, $\Cc^{(1)}=\{[0]\}$,
$\Cc^{(2)}=\{[01],[10]\}$, and
$\Cc^{(3)}=\{[102],[120]\}$. Hence, $\dmin(\Cc^{(3)})=2$.

\subsection{Number of Size-Preserving Extensions Needed to Increase Minimum Distance}
A length-$n$ code with minimum distance at least $d$ is called an $[n,d]$ code. A singleton head set preserves code size, whereas a larger head set increases it. We therefore measure the minimum number of future singleton-head extensions required to turn an intermediate code into an $[n,d]$ code. This number records the part of the remaining construction that cannot contribute to cardinality.

Let $d\ge1$ be an integer and let $\Cc_0$ be a code in $\Sc_N$. For $j\ge0$, choose a nonempty head set $S_j\subset[N+j+1]$ and set $\Cc_{j+1}:=\Cc_j^{S_j}$. Define $c_1(\Cc_0;d)$ as the minimum number of extension steps with singleton head sets among all finite extension sequences that produce a code of minimum distance at least $d$.
For a fixed number $k$ of extension steps, let $c_1^{(k)}(\Cc_0;d)$ be the minimum number of size-preserving steps among all length-$k$ extension sequences whose final code has minimum distance at least $d$. 
Formally,
\begin{align}
&c_1(\Cc_0; d) \defeq \min_{k \ge 0} c_1^{(k)}(\Cc_0; d) \label{151909_28Apr24}
\\
&c_1^{(k)}(\Cc_0; d) \defeq
\min_{\substack{S_0,\ldots,S_{k-1}\\ \dmin(\Cc_k)\ge d}}
\#\{l\in[k]:|S_l|=1\}
\end{align}
Thus $c_1(\Cc_0;d)$ is not the number of all future steps; it counts only the steps at which the encoder has no choice of head and hence gains no new codewords.
If no length-$k$ sequence satisfies the distance requirement, define $c_1^{(k)}(\Cc_0;d):=\infty$.
For $k=0$, the empty extension sequence has cost zero when $\dmin(\Cc_0)\ge d$; otherwise it is infeasible. In particular,
$c_1(\Cc_0;d)=0$ whenever $\Cc_0$ already has minimum distance at least $d$.
The value $c_1(\Cc_0;d)$ is finite. For each intermediate code $\Cc_j$, define the total amount by which its pairwise distances fall below $d$ as
\[
\Delta_j:=\sum_{\{\piU,\sigmaU\}\in\binom{\Cc_j}{2}}
\max\{d-d_\infty(\piU,\sigmaU),0\}.
\]
Suppose that $\piU,\sigmaU\in\Cc_j$ have distance below $d$. Choose a coordinate $i$ such that $|\pi_i-\sigma_i|=d_\infty(\piU,\sigmaU)$, choose $s\in\Ic(\pi_i,\sigma_i)$, and extend the code using the singleton head set $\{s\}$. Lemma~\ref{lem:phi_kakudai} increases the selected pair's distance by one and does not decrease the distance of any other pair. By Lemma~\ref{152042_1Oct23}, this extension gives a one-to-one correspondence between the codeword pairs before and after extension. Hence the selected pair's contribution to $\Delta_j$ decreases by one, while no other contribution increases, so $\Delta_{j+1}\le\Delta_j-1$. Because $\Delta_0$ is a finite nonnegative integer, after at most $\Delta_0$ such extensions every pair has distance at least $d$. Thus a finite extension sequence satisfying the required minimum distance always exists.

For an intermediate code lying on a completed length-$n$ REP construction, the remaining $n-k$ stages themselves give an immediate upper bound on this required number.
\begin{lem}[Upper bound on $c_1$]\label{165523_23May24}
Let $n>d\ge1$, and let $\Cc^{(n)}$ be an REP code with $\dmin(\Cc^{(n)})\ge d$. For every $1\le k\le n$,
 \begin{align}
 c_1(\Cc^{(k)};d) \le n - k{.} \label{031411_8May24}
 \end{align}
\end{lem}
\begin{IEEEproof}
The fixed sequence of head sets $S^{(k)},\ldots,S^{(n-1)}$ extends $\Cc^{(k)}$ to $\Cc^{(n)}$, whose minimum distance is at least $d$. It is therefore a feasible sequence in the definition of $c_1(\Cc^{(k)};d)$. This sequence has $n-k$ steps and hence contains at most $n-k$ singleton head sets, which proves \eqref{031411_8May24}.
\end{IEEEproof}
{\begin{example}
Let $n=6$, $d=2$, and consider the REP code constructed from
\begin{align*}
S^{(0)} &= \{0\}, & S^{(1)} &= \{0\},\\
S^{(2)} &= \{0,2\}, & S^{(3)} &= \{0,2\},\\
S^{(4)} &= \{0,2,4\}, & S^{(5)} &= \{0,2,4\}.
\end{align*}
This yields a code $\Cc^{(6)}$ with $\dmin(\Cc^{(6)}) \ge 2$.
For $k=4$, the displayed extension from $\Cc^{(4)}$ to $\Cc^{(6)}$ contains no singleton head sets. Hence
$c_1(\Cc^{(4)}; 2) = 0 \le 6 - 4 = 2$,
as required by Lemma~\ref{165523_23May24}.
\end{example}
}

The effect of a size-increasing step depends on how closely its heads are placed. For a head set $S\subset[n+1]$, let $\delta_d(S)$ denote the total amount by which its consecutive spacings fall below $d$. Set $\delta_d(S)=0$ when $|S|=1$. When $S=\{s_1<\cdots<s_q\}$ with $q\ge2$, define
\begin{align}
  \delta_d(S) \defeq \sum_{\substack{1\le j<q\\s_{j+1}-s_j<d}} (d - (s_{j+1}-s_j)). \label{151901_28Apr24}
\end{align}
Each term is the number of additional distance units needed to raise one short consecutive spacing to $d$; $\delta_d(S)$ sums these requirements over all consecutive heads.
For every code $\Cc_0\subset\Sc_n$, Theorem~\ref{231412_20May24} gives $c_1(\Cc_0^S;d)\ge \delta_d(S)$.

The spacing shortfall also limits the size of the head set. Allowing total shortfall at most $c$ increases the usual packing bound by at most $c/d$. Theorem~\ref{230931_30Oct23} is recovered after replacing $n$ below by $n+1$ and setting $c=0$.
\begin{lem}\label{234753_24Apr24} 
Let $n,d\ge1$ be integers. If $S\subset[n]$ and $c\ge \delta_d(S)$, then
\begin{align}
 |S| \le \frac{n-1+c}{d} + 1.\label{132307_12Jul24}
\end{align}
\end{lem}
\begin{IEEEproof}
Write $S=\{s_1<\cdots<s_{|S|}\}$ and let
$J:=\{1,\ldots,|S|-1\}$,
$\JU:=\{j\in J:s_{j+1}-s_j<d\}$, and
$\JO:=J\setminus\JU$. Then
 \begin{align}
 n &\ge 1 + \sum_{j \in J} (s_{j+1}-s_j)
 \\&= 1 + \sum_{j \in \JU} (s_{j+1}-s_j) + \sum_{j \in \JO} (s_{j+1}-s_j)
 \\&\ge 1 + |\JU|d - c + |\JO|d
 \\&= 1 - c + (|S|-1)d.
 \end{align}
The first inequality follows from $s_{|S|}-s_1\le n-1$. For the second, use
$c\ge\sum_{j\in\JU}(d-(s_{j+1}-s_j))$ and
$s_{j+1}-s_j\ge d$ for $j\in\JO$.
Rearranging proves \eqref{132307_12Jul24}.
\end{IEEEproof}
{
\begin{example}
Let $n=10$, $d=3$, and $S=\{0,2,5,6,9\}\subset[10]$. The consecutive gaps are $(2,3,1,3)$, so $\delta_3(S)=1+2=3$. Lemma~\ref{234753_24Apr24} gives
\[
|S| \le \left\lfloor \frac{n - 1 + \delta_3(S)}{3} \right\rfloor + 1 = \left\lfloor \frac{12}{3} \right\rfloor + 1 = 4 + 1 = 5.
\]
Thus equality holds.
\end{example}
}

We next relate the required number of singleton steps before and after one extension. One new extension can reduce that requirement by at most one, and a reduction by exactly one is possible only when the extension itself uses a singleton head set.
\begin{teiri}\label{153614_27May24} 
For a code $\Cc_0\subset\Sc_n$, a head set $S_0\subset[n+1]$, and $d\ge1$, let $\Cc_1=\Cc_0^{S_0}$. Then
\begin{enumerate*}[label=\roman*), series = tobecont, itemjoin = \quad]
\item \label{161656_27May24} $c_1(\Cc_0;d) \le c_1(\Cc_1;d) + 1$;
\item \label{161706_27May24} $c_1(\Cc_0;d) = c_1(\Cc_1;d) + 1$ implies $|S_0| = 1$.
\end{enumerate*}
\end{teiri}
\begin{IEEEproof}
Take an extension sequence from $\Cc_1$ attaining $c_1(\Cc_1;d)$ and prepend the extension by $S_0$. The resulting sequence is feasible for $\Cc_0$ and contains at most one additional singleton head set. This proves part~\ref{161656_27May24}.

For part~\ref{161706_27May24}, suppose $|S_0|\ne1$. Prepending $S_0$ to the same sequence adds no singleton head set, so $c_1(\Cc_0;d)\le c_1(\Cc_1;d)$. This contradicts $c_1(\Cc_0;d)=c_1(\Cc_1;d)+1$.
\end{IEEEproof}
{\begin{example}
Let $\Cc_0=\{[01],[10]\}\subset\Sc_2$, $d=2$, and $S_0=\{1\}$. Then
$\Cc_1=\Cc_0^{S_0}=\{[102],[120]\}$ and $\dmin(\Cc_1)=2$, so $c_1(\Cc_1;2)=0$. The maximum interval of the pair $([01],[10])$ is $(0,1]$; an extension with a non-singleton head set cannot increase the corresponding maximum-interval length by Lemma~\ref{152448_18May24}. Hence an extension with a singleton head set is necessary, while the extension by $S_0$ shows that one is sufficient. Thus $c_1(\Cc_0;2)=1=c_1(\Cc_1;2)+1$ and $|S_0|=1$.
\end{example}
}

For a codeword pair $\Psf=(\piU,\sigmaU)$, write $\Psf^s=(\piU^s,\sigmaU^s)$. If $I$ is a maximum interval of $\Psf$, Lemma~\ref{060127_4Mar24} gives
$\dinf(\Psf^s)=\dinf(\Psf)+\I[s\in I]$ and hence $|I^s|=|I|+\I[s\in I]$.

Several short spacings in one head set create several codeword pairs that must all reach distance $d$. Their pairwise-disjoint maximum intervals can be followed through every later extension. A singleton-head step can increase at most one of them by one, while a step with several heads can be represented by pairs whose tracked intervals do not increase.
\begin{lem}\label{152448_18May24} 
Let $\Cc$ be a length-$n$ code containing codeword pairs $\Psf_1,\ldots,\Psf_k$. Suppose their maximum intervals $I_1,\ldots,I_k$ are pairwise disjoint. For any nonempty $S\subset[n+1]$, the extended code $\Cc^S$ contains pairs $\Qsf_1,\ldots,\Qsf_k$ with pairwise-disjoint maximum intervals $J_1,\ldots,J_k$ such that:
\begin{enumerate}[label=\arabic*)]
    \item if $|S|=1$, then, after relabeling if necessary, $|J_1|\le|I_1|+1$ and $|J_i|\le|I_i|$ for $i\ne1$;
    \item if $|S|\ge2$, then $|J_i|\le|I_i|$ for every $i$.
\end{enumerate}
\end{lem}

\begin{IEEEproof}\label{225701_20May24}
First suppose $S=\{s\}$ and set $\Qsf_i:=\Psf_i^s$. By Lemma~\ref{060127_4Mar24}, $\Qsf_i$ has maximum interval $J_i:=I_i^s$ with
$|J_i|=|I_i|+\I[s\in I_i]$. Because the $I_i$ are pairwise disjoint, at most one contains $s$. The intervals $J_i$ remain pairwise disjoint by Lemma~\ref{153103_22May24}\ref{045307_11Apr25}, proving the first claim.

Now suppose $|S|\ge2$ and choose distinct $s,t\in S$. After interchanging $s$ and $t$ and relabeling the intervals if necessary, one of the following cases applies.

\begin{enumerate}[label=\alph*), leftmargin=2em]
\item \label{155653_10Apr25}
\textit{$s$ lies in no $I_i$.}
Set $\Qsf_i:=\Psf_i^s$ for every $i$. Lemma~\ref{060127_4Mar24} gives the maximum intervals $J_i=I_i^s$ with $|J_i|=|I_i|$, and Lemma~\ref{153103_22May24}\ref{045307_11Apr25} preserves their pairwise disjointness.

\item 
\textit{$s$ and $t$ lie in the same interval.}
Suppose $s,t\in I_1$. Fix $\piU\in\Cc$, set $\Qsf_1:=(\piU^s,\piU^t)$, and set $\Qsf_i:=\Psf_i^s$ for $i\ge2$. By Lemma~\ref{000804_8Dec23}, $\Qsf_1$ has maximum interval
$J_1=\Ic(s,t)\subsetneq I_1$, so $|J_1|<|I_1|$. For $i\ge2$, case~\ref{155653_10Apr25} gives $J_i=I_i^s$ and $|J_i|=|I_i|$. Lemma~\ref{153103_22May24}\ref{045307_11Apr25} shows that $I_1^s$ is disjoint from every $I_i^s$; since $J_1\subset I_1^s$, all $J_i$ are pairwise disjoint.

\item \textit{$s$ and $t$ lie in different intervals.}
Suppose $s\in I_1$ and $t\in I_2$. Set $\Qsf_1:=\Psf_1^t$, $\Qsf_2:=\Psf_2^s$, and $\Qsf_i:=\Psf_i^s$ for $i\ge3$. Because each pair is extended by a head outside its maximum interval, Lemma~\ref{060127_4Mar24} gives $|J_i|=|I_i|$ for every $i$. Lemma~\ref{153103_22May24}\ref{045314_11Apr25} shows that $J_1=I_1^t$ and $J_2=I_2^s$ are disjoint. The intervals $J_i=I_i^s$ for $i\ge2$ are pairwise disjoint by case~\ref{155653_10Apr25}, and Lemma~\ref{153103_22May24}\ref{045321_11Apr25} shows that $J_1$ is disjoint from each of them.
\end{enumerate}
Thus $|J_i|\le|I_i|$ in every case.
\end{IEEEproof}

{\begin{example}
Let $\Cc$ contain $\piU=[01234]$, $\sigmaU=[01324]$, and $\tauU=[01243]$. The pairs $\Psf_1=(\piU,\sigmaU)$ and $\Psf_2=(\piU,\tauU)$ have disjoint maximum intervals $I_1=(2,3]=\{3\}$ and $I_2=(3,4]=\{4\}$.

For $S=\{3\}$, take
\begin{align}
\Qsf_1&=(\piU^3,\sigmaU^3)=([301245],[301425]),\\
\Qsf_2&=(\piU^3,\tauU^3)=([301245],[301254]).
\end{align}
Their maximum intervals are
\begin{align}
J_1&=(2,4]=\{3,4\}, & J_2&=(4,5]=\{5\}.
\end{align}
For $S=\{1,4\}$, take
\begin{align}
\Qsf_1&=(\piU^1,\sigmaU^1)=([102345],[102435]),\\
\Qsf_2&=(\piU^1,\tauU^1)=([102345],[102354]).
\end{align}
The corresponding maximum intervals are
\begin{align}
J_1&=(3,4]=\{4\}, & J_2&=(4,5]=\{5\}.
\end{align}
\end{example}
}

The spacing shortfall created by one head set now gives the required lower bound: at least that many later singleton-head extensions are needed before the code can attain distance $d$.
\begin{teiri}\label{231412_20May24} 
For an integer $d\ge1$, a code $\Cc_0\subset\Sc_n$, and a head set $S_0\subset[n+1]$, let $\Cc_1:=\Cc_0^{S_0}$. Then $c_1(\Cc_1;d)\ge \delta_d(S_0)$.
\end{teiri}
\begin{IEEEproof}\label{231611_20May24}
Take any finite extension sequence $S_1,\ldots,S_{m-1}$. For $j=1,\ldots,m-1$, let $S_j\subset[n+j+1]$ and $\Cc_{j+1}:=\Cc_j^{S_j}$. Assume $\dmin(\Cc_m)\ge d$. Thus $\Cc_j$ has length $n+j$. It suffices to show that the sequence contains at least $\delta_d(S_0)$ singleton head sets.
If $|S_0|=1$, then $\delta_d(S_0)=0$ and the result is immediate. Assume $|S_0|\ge2$. Let the elements of $S_0$ be $s_1 < \cdots < s_{k+1}$, where $k := |S_0| - 1$.
Choose $\piU\in\Cc_0$ and denote the $k$ pairs $(\piU^{s_i},\piU^{s_{i+1}})$ in $\Cc_1$ by $\Psf_i^{(1)}$.
Each pair $\Psf_i^{(1)}$ has maximum interval $I_i^{(1)} := {\Ic}(s_i, s_{i+1})$, and these intervals are mutually disjoint.
Apply Lemma~\ref{152448_18May24} successively to $\Cc_r$ and $S_r$ for $r=1,\ldots,m-1$. This yields pairs $\Psf_1^{(m)},\ldots,\Psf_k^{(m)}$ in $\Cc_m$ whose maximum intervals remain pairwise disjoint and satisfy the lemma's length bounds.
Since $\dmin(\Cc_m) \ge d$, the maximum interval of every pair $\Psf_i^{(m)}$ has length at least $d$.
Lemma~\ref{152448_18May24} permits an interval to increase only under an extension with a singleton head set, and then by at most one; moreover, at most one of the $k$ intervals increases at a given step. Raising all initial interval lengths $s_{i+1}-s_i<d$ to at least $d$ therefore requires at least
\[
\sum_{\substack{i\\s_{i+1}-s_i<d}}
\bigl(d-(s_{i+1}-s_i)\bigr)=\delta_d(S_0)
\]
singleton extensions.
\end{IEEEproof}

{\begin{example}
Let $n=4$, $d=3$, $\Cc_0=\{[0123]\}\subset\Sc_4$, and $S_0=\{0,2,3\}\subset[5]$. Since $\delta_3(S_0)=1+2=3$, Theorem~\ref{231412_20May24} gives
\[
c_1(\Cc_0^{S_0};3)\ge3.
\]
Thus at least three subsequent extension steps with singleton head sets are required to reach minimum distance at least~3.
\end{example}
}

\subsection{Maximum-Size REP Codes}
\label{161538_26Aug24}

The preceding results quantify both sides of the tradeoff. A large head set multiplies the code size, but its spacing shortfall forces later singleton steps whose multiplication factor is one. We now combine these facts over the entire extension sequence, derive an upper bound, and attain it with equally spaced heads. All maxima in this subsection are taken over REP codes, and $n>d\ge1$.

The proof uses the following elementary statement about the sequence of remaining singleton-step requirements. It bounds how many stages can contribute a nontrivial multiplication factor and how large those factors can be.
\begin{lem}\label{021210_24Jul24}
Let $c^{(0)},\ldots,c^{(n)}$ be nonnegative integers satisfying
$c^{(k)}\le c^{(k+1)}+1$ for $0\le k<n$ and
$c^{(k)}\le n-k$ for $1\le k\le n$. Define
\[
K:=\{0\le k<n:c^{(k)}\le c^{(k+1)}\}.
\]
If $K$ is nonempty and $K=\{k_1<\cdots<k_m\}$, then, for $i=1,\ldots,m-1$,
\begin{align}
 k_i + c^{(k_i+1)} \le n-1-(m-i).\label{014925_24Jul24}
\end{align}
\end{lem}
\begin{IEEEproof}
We first show that, for $i=1,\ldots,m-1$,
\begin{align}
  k_i + c^{(k_i+1)} < k_{i+1} + c^{(k_{i+1}+1)}. \label{172920_9May24}
\end{align}
Since $k_{i+1}\in K$, we have $c^{(k_{i+1})}\le c^{(k_{i+1}+1)}$. There are two cases.

\begin{enumerate}
  \item If $k_i+1=k_{i+1}$, then $c^{(k_i+1)}=c^{(k_{i+1})}$, and hence
  \[
    k_i + c^{(k_i+1)} = k_{i+1} - 1 + c^{(k_{i+1})}
    \le k_{i+1} + c^{(k_{i+1}+1)} - 1.
  \]

  \item If $k_i+1<k_{i+1}$, then $k_i+1,\ldots,k_{i+1}-1$ do not belong to $K$. The first hypothesis and integrality therefore give
  \[
    c^{(k)} = c^{(k+1)} + 1
    \quad\text{for }k_i+1\le k<k_{i+1}.
  \]
  Telescoping yields
  \[
    c^{(k_i+1)} - c^{(k_{i+1})} = k_{i+1} - k_i - 1,
  \]
  which implies
  \[
    k_i + c^{(k_i+1)}
    = k_{i+1} - 1 + c^{(k_{i+1})}
    \le k_{i+1} + c^{(k_{i+1}+1)} - 1.
  \]
\end{enumerate}
Thus \eqref{172920_9May24} holds.

The second hypothesis gives
\begin{align}
  k_m + c^{(k_m+1)} \le n - 1. \label{060246_11Apr25}
\end{align}
Applying \eqref{172920_9May24} repeatedly, starting from \eqref{060246_11Apr25}, proves \eqref{014925_24Jul24}.
\end{IEEEproof}

The resulting upper bound has one factor for each possible code length. It states that no REP construction can gain more total cardinality than the product obtained by placing heads $d$ units apart at every stage.
\begin{teiri}\label{165623_23May24}
For integers $n>d\ge1$, every $[n,d]$ REP code $\Cc^{(n)}$ satisfies
$|\Cc^{(n)}|\le\prod_{j=0}^{n-1}(\lfloor j/d\rfloor+1)$.
\end{teiri}
\begin{IEEEproof}
To simplify notation, write $c^{(k)} := c_1(\Cc^{(k)}; d)$ for $0\le k\le n$. Since $\Cc^{(0)}$ contains only the empty permutation and $\Cc^{(n)}$ has minimum distance at least $d$, we have $c^{(0)}=c^{(n)}=0$.
Let $K$ and $K^\mathrm{c}$ index the nondecreasing and decreasing steps, respectively:
$K \defeq \{0 \le k <n : c^{(k)} \le c^{(k+1)}\}$, 
$K^\mathrm{c} \defeq \{0 \le k <n : c^{(k)} > c^{(k+1)}\}$.
In particular, $0\in K$, so $K$ is nonempty.
For $k=0$, the inequality $c^{(0)}\le c^{(1)}+1$ follows from $c^{(0)}=0$. For $1\le k<n$, Theorem~\ref{153614_27May24} gives $c^{(k)}\le c^{(k+1)}+1$, and Lemma~\ref{165523_23May24} gives $c^{(k)}\le n-k$ for $1\le k\le n$. Thus the hypotheses of Lemma~\ref{021210_24Jul24} hold.
For $k \in K^\mathrm{c}$, Theorem~\ref{153614_27May24} gives
$c^{(k)} = c^{(k+1)} + 1$ and $|S^{(k)}| = 1$.
Hence,
\[
|\Cc^{(n)}|=\prod_{k=0}^{n-1} |S^{(k)}| 
= \prod_{k \in K} |S^{(k)}|.
\]
Writing $K=\{k_1<\cdots<k_{|K|}\}$ and applying Theorem~\ref{231412_20May24} and Lemma~\ref{234753_24Apr24}, we obtain
\begin{align}
\prod_{k \in K} |S^{(k)}| &\le \prod_{k \in K} \left(\left\lfloor \frac{k + c^{(k+1)}}{d}\right\rfloor + 1\right) \label{020938_27Apr24} \\
&= \prod_{i=1}^{|K|} \left(\left\lfloor \frac{k_i + c^{(k_i+1)}}{d}\right\rfloor + 1\right).
\end{align}
For $|K| = 1$, Lemma~\ref{165523_23May24} gives $k_1 + c^{(k_1+1)} \le n-1$. Hence, the single factor in \eqref{020938_27Apr24} is at most $\lfloor(n-1)/d\rfloor+1$, which is the $j=n-1$ factor in the claimed product; all other factors are at least one. It remains to consider $|K| \ge 2$. Then
\begin{align}
\prod_{i=1}^{|K|} \left(\left\lfloor \frac{k_i + c^{(k_i+1)}}{d}\right\rfloor + 1\right) &{\le} \prod_{i=1}^{|K|} \left(\left\lfloor \frac{n-1-(|K|-i)}{d}\right\rfloor + 1\right).
\end{align}
This inequality follows from Lemma~\ref{021210_24Jul24}. Finally,
\begin{align}
 \prod_{i=1}^{|K|} \left(\left\lfloor \frac{n-1-(|K|-i)}{d}\right\rfloor + 1\right)
 &= \prod_{j=n-|K|}^{n-1} (\lfloor j/d\rfloor+1) \notag\\
 &\le \prod_{j=0}^{n-1} (\lfloor j/d\rfloor+1).
\end{align}
\end{IEEEproof}
{\begin{example}\label{225521_13Apr25}
Let $n = 6$ and $d = 2$. Theorem~\ref{165623_23May24} gives an upper bound on the size of any REP code with minimum distance at least $d$:
\[
|\Cc^{(6)}| \le \prod_{j=0}^{5} \left( \left\lfloor \frac{j}{2} \right\rfloor + 1 \right)
= 1\cdot 1\cdot 2\cdot 2\cdot 3\cdot 3 = 36.
\]
This upper bound is tight, since it is achieved by the REP code constructed as follows:
\begin{align*}
S^{(0)} &= \{0\}, & S^{(1)} &= \{0\},\\
S^{(2)} &= \{0,2\}, & S^{(3)} &= \{0,2\},\\
S^{(4)} &= \{0,2,4\}, & S^{(5)} &= \{0,2,4\}.
\end{align*}
Each $S^{(j)}$ satisfies $\dmin(S^{(j)}) \ge 2$, and the code size is
\[
|\Cc^{(6)}| = 1 \cdot 1 \cdot 2 \cdot 2 \cdot 3 \cdot 3 = 36.
\]
\end{example}
}

The bound is attainable: choosing all available multiples of $d$ as heads at each stage meets the spacing requirement without creating any future shortfall.
\begin{teiri}\label{223851_4Jun24}
For integers $n>d\ge1$, there exists an $[n,d]$ REP code $\Cc^{(n)}$ of size $|\Cc^{(n)}| = \prod_{j=0}^{n-1}(\lfloor j/d\rfloor+1)$.
\end{teiri}
\begin{IEEEproof}
For $j=0,\ldots,n-1$, choose
\[
S^{(j)}:=\{0,d,2d,\ldots,\lfloor j/d\rfloor d\}\subset[j+1].
\]
Then $|S^{(j)}|=\lfloor j/d\rfloor+1$, while $\dmin(S^{(j)})=\infty$ for $j<d$ and $\dmin(S^{(j)})=d$ for $j\ge d$. Repeated application of Theorems~\ref{205810_6Jul23} and~\ref{163227_14May23} yields
\[
|\Cc^{(n)}|=\prod_{j=0}^{n-1}|S^{(j)}|
=\prod_{j=0}^{n-1}(\lfloor j/d\rfloor+1)
\]
and $\dmin(\Cc^{(n)})\ge d$.
\end{IEEEproof}
By \eqref{dpgp_product}, the maximum REP-code size therefore equals the cardinality of the corresponding DPGP code.
This equality answers the extremal question within the REP family. It compares cardinalities, not the complete geometry of the two codes.

Equality of cardinalities does not imply that the two codes have the same pairwise distances. For $n=4$ and $d=2$, the construction in Theorem~\ref{223851_4Jun24} gives $\Cc^{(4)}=\{[0123],[2013],[0312],[2301]\}$, whose six pairwise distances are $2,2,2,2,3,3$. The corresponding DPGP code $\{[0123],[0321],[2103],[2301]\}$ has all six pairwise distances equal to $2$. Hence no bijection between these two codes preserves every pairwise Chebyshev distance.

\section{Encoding and Decoding Algorithms}
The extremal argument treats an REP code as a sequence of head choices. The same representation is also a message description. We first translate the defining recursion directly into an encoder, then remove its repeated relabeling by interpreting the reversed head sequence as ranks among unused symbols. The resulting coordinate-order representation also identifies the candidates that must be compared during bounded-distance decoding.

Consider the REP code $\Cc^{(n)}=\<S^{(0)},\ldots,S^{(n-1)}\>$. Its encoder input is a head sequence
$(s^{(0)},\ldots,s^{(n-1)})\in S^{(0)}\times\cdots\times S^{(n-1)}$.\footnote{Equivalently, list $S^{(j)}=\{a_0^{(j)}<\cdots<a_{|S^{(j)}|-1}^{(j)}\}$. A message consists of mixed-radix digits $x_j\in[|S^{(j)}|]$, with $s^{(j)}=a_{x_j}^{(j)}$.}
\subsection{Direct Recursive Encoding Algorithm}
\label{182618_25May23}
The direct encoder is the computational form of the definition. It makes correctness immediate and supplies a reference implementation, but it repeatedly updates previously created symbols and therefore requires quadratic time.

The direct recursive encoder applies the defining extension operation at each step. Starting from $\piU^{(0)}=\varepsilonU$, it computes
\[
\piU^{(j)}=\phi(\piU^{(j-1)};s^{(j-1)}), \qquad j=1,\ldots,n.
\]
Repeated application of Lemma~\ref{152042_1Oct23} shows that distinct head sequences produce distinct codewords. Thus this procedure generates every codeword of $\Cc^{(n)}$ exactly once. We write its output as $\piU^{(n)}=\Cc^{(n)}(\sU)$.

 \begin{figure*}[!t]
 \begin{center}
{\scriptsize  
\begin{tabular}{
@{\hspace{0mm}}l||@{\hspace{2mm}}c@{\hspace{0mm}}c@{\hspace{0mm}}c@{\hspace{0mm}}c@{\hspace{0mm}}c@{\hspace{0mm}}c@{\hspace{0mm}}c@{\hspace{0mm}}c@{\hspace{0mm}}c@{\hspace{0mm}}c@{\hspace{0mm}}c@{\hspace{0mm}}c@{\hspace{0mm}}c@{\hspace{0mm}}c@{\hspace{0mm}}c@{\hspace{0mm}}c@{\hspace{0mm}}c@{\hspace{0mm}}c@{\hspace{0mm}}c@{\hspace{0mm}}c@{\hspace{0mm}}c@{\hspace{0mm}}c@{\hspace{0mm}}c@{\hspace{0mm}}c@{\hspace{0mm}}c@{\hspace{0mm}}c@{\hspace{0mm}}c@{\hspace{0mm}}c@{\hspace{0mm}}c@{\hspace{0mm}}
}
   $\piU^{(8)}$&$s^{(7)}=$&$\pi_0^{(8)}$ && $\pi_1^{(8)}$ && $\pi_2^{(8)}$ && $\pi_3^{(8)}$ && $\pi_4^{(8)}$ && $\pi_5^{(8)}$ && $\pi_6^{(8)}$ && $\pi_7^{(8)}$
\\ $\uparrow_7$&&& $\nearrow\hspace{-3mm}{}^7$&&$\nearrow\hspace{-3mm}{}^7$&&$\nearrow\hspace{-3mm}{}^7$&&$\nearrow\hspace{-3mm}{}^7$&&$\nearrow\hspace{-3mm}{}^7$&&$\nearrow\hspace{-3mm}{}^7$&&$\nearrow\hspace{-3mm}{}^7$&&
\\ $\piU^{(7)}$&$s^{(6)}=$&$\pi_0^{(7)}$ && $\pi_1^{(7)}$ && $\pi_2^{(7)}$ && $\pi_3^{(7)}$ && $\pi_4^{(7)}$ && $\pi_5^{(7)}$ && $\pi_6^{(7)}$ && 
\\ $\uparrow_6$&&& $\nearrow\hspace{-3mm}{}^6$&&$\nearrow\hspace{-3mm}{}^6$&&$\nearrow\hspace{-3mm}{}^6$&&$\nearrow\hspace{-3mm}{}^6$&&$\nearrow\hspace{-3mm}{}^6$&&$\nearrow\hspace{-3mm}{}^6$&&
\\ $\piU^{(6)}$&$s^{(5)}=$&$\pi_0^{(6)}$ && $\pi_1^{(6)}$ && $\pi_2^{(6)}$ && $\pi_3^{(6)}$ && $\pi_4^{(6)}$ && $\pi_5^{(6)}$ && 
\\ $\uparrow_5$&&& $\nearrow\hspace{-3mm}{}^5$&&$\nearrow\hspace{-3mm}{}^5$&&$\nearrow\hspace{-3mm}{}^5$&&$\nearrow\hspace{-3mm}{}^5$&&$\nearrow\hspace{-3mm}{}^5$&&
\\ $\piU^{(5)}$&$s^{(4)}=$&$\pi_0^{(5)}$ && $\pi_1^{(5)}$ && $\pi_2^{(5)}$ && $\pi_3^{(5)}$ && $\pi_4^{(5)}$ && 
\\ $\uparrow_4$&&& $\nearrow\hspace{-3mm}{}^4$&&$\nearrow\hspace{-3mm}{}^4$&&$\nearrow\hspace{-3mm}{}^4$&&$\nearrow\hspace{-3mm}{}^4$&&
\\ $\piU^{(4)}$&$s^{(3)}=$&$\pi_0^{(4)}$ && $\pi_1^{(4)}$ && $\pi_2^{(4)}$ && $\pi_3^{(4)}$ &&
\\ $\uparrow_3$&&& $\nearrow\hspace{-3mm}{}^3$&&$\nearrow\hspace{-3mm}{}^3$&&$\nearrow\hspace{-3mm}{}^3$&&
\\ $\piU^{(3)}$&$s^{(2)}=$&$\pi_0^{(3)}$ && $\pi_1^{(3)}$ && $\pi_2^{(3)}$ && 
\\ $\uparrow_2$&&& $\nearrow\hspace{-3mm}{}^2$&&$\nearrow\hspace{-3mm}{}^2$&&
\\ $\piU^{(2)}$&$s^{(1)}=$&$\pi_0^{(2)}$ && $\pi_1^{(2)}$ && 
\\ $\uparrow_1$&&& $\nearrow\hspace{-3mm}{}^1$&&
\\ $\piU^{(1)}$&$s^{(0)}=$&$\pi_0^{(1)}$ && 
 \end{tabular}
} 
\end{center}
\caption{Dependencies among variables in the direct recursive encoding algorithm for $n=8$. We write $a \xrightarrow{j} b$ when $b = \phi_{s^{(j)}}(a)$.}
\label{125759_3Aug23}
\end{figure*}

\begin{figure}[!t]
    \begin{algorithm}[H]
        \caption{Direct Recursive Encoder for $\Cc^{(n)}$}
        \label{natural_encode}
        \begin{algorithmic}[1]
        \renewcommand{\algorithmicrequire}{\textbf{Input:}}
        \renewcommand{\algorithmicensure}{\textbf{Output:}}
        \REQUIRE $ (s^{(0)} , \dotsc, s^{(n-1)})\in S^{(0)}\times\cdots\times S^{(n-1)}$\\
        \ENSURE  $(\pi_0^{(n)} , \dotsc, \pi_{n-1}^{(n)})\in \Cc^{(n)}$
        \FOR {$j:= 1$ to $n$}
            \STATE $\pi^{(j)} _0 := s^{(j-1)}$
		 \ENDFOR
		 \FOR {$k := 1$ to $n-1$}
		 \FOR {$j:= k+1$ to $n$}\label{141220_6Aug23}
                \STATE $\pi_k^{(j)}:=\phi_{s^{(j-1)}}\left(\pi_{k-1}^{(j-1)}\right)$\label{141312_6Aug23}
            \ENDFOR\label{141314_6Aug23}
        \ENDFOR
        \RETURN $(\pi_0^{(n)},\ldots,\pi_{n-1}^{(n)})$
        \end{algorithmic}
    \end{algorithm}
\end{figure}
Algorithm~\ref{natural_encode} gives a componentwise implementation, and Fig.~\ref{125759_3Aug23} shows its variable dependencies for $n=8$. This direct implementation requires $O(n^2)$ operations.
\subsection{Sequential Encoding Algorithm}
\label{182606_25May23}
The final permutation can instead be produced from left to right without constructing every intermediate permutation. The head used at the last extension is the first output symbol; each earlier head gives the rank of a later output symbol among those still unused.

A sequential encoder produces $\pi_0^{(n)},\ldots,\pi_{n-1}^{(n)}$ in coordinate order while storing the symbols already produced. The dependencies in Fig.~\ref{125759_3Aug23} permit this order, but direct evaluation still requires $O(n^2)$ operations. Algorithm~\ref{214627_21Aug24} reduces the cost to $O(n\log n)$ by treating $s^{(n-1)},\ldots,s^{(0)}$ as Lehmer-code digits: each digit specifies the zero-based rank of the next output symbol among the unused symbols. The constraints $s^{(j)}\in S^{(j)}$ restrict the possible digits, and standard permutation unranking produces the codeword~\cite{lehmer_unranking}.

For a set $A\subset[n]$, define its image by $\phi'_s(A):=\{\phi_s(a):a\in A\}$ and its augmented image by $\phi_s(A):=\phi'_s(A)\cup\{s\}$.

For a finite set $A=\{a_0<\cdots<a_{|A|-1}\}$ and $r\in[|A|]$, define $\operatorname{select}_r(A):=a_r$.
Thus $\operatorname{select}_r(A)$ is simply the element of zero-based rank $r$ in the ordered set $A$. The first compatibility lemma says that extension preserves this rank.
\begin{lem}\label{181528_2Aug23}
For $A\subset[n]$, $s\in[n+1]$, and $r\in[|A|]$,
\begin{align}
\phi_s(\operatorname{select}_r(A))=\operatorname{select}_r\bigl(\phi_s'(A)\bigr). \label{223340_2Aug23}
\end{align}
\end{lem}
\begin{IEEEproof}
List the elements of $A$ as $a_0<\cdots<a_{|A|-1}$. Because $\phi_s$ is strictly increasing,
$\phi_s(a_0)<\cdots<\phi_s(a_{|A|-1})$. Both sides of \eqref{223340_2Aug23} therefore equal $\phi_s(a_r)$.
\end{IEEEproof}

For $X\subset[n]$, write $\overline{X}^{[n]}:=[n]\setminus X$.
The second compatibility lemma says that relabeling all unused symbols gives exactly the unused symbols after extension.
\begin{lem}\label{224537_2Aug23}
For $A\subset[n-1]$ and $s\in[n]$,
$\phi_s'(\overline{A}^{[n-1]}) = \overline{\phi_s(A)}^{[n]}$.
\end{lem}
\begin{IEEEproof}
The sets $A$ and $\overline{A}^{[n-1]}$ partition $[n-1]$. Since $\phi'_s$ is a bijection from $[n-1]$ to $[n]\setminus\{s\}$, the three sets $\phi'_s(A)$, $\phi'_s(\overline{A}^{[n-1]})$, and $\{s\}$ are pairwise disjoint and have union $[n]$. Taking the complement of $\phi_s(A)=\phi'_s(A)\cup\{s\}$ in $[n]$ gives
$\overline{\phi_s(A)}^{[n]}=\phi'_s(\overline{A}^{[n-1]})$.
\end{IEEEproof}

These two compatibility statements prove that selecting by rank from the unused symbols produces the same coordinate as the defining recursion.
\begin{teiri}\label{153741_1Aug23}
For $n\ge1$ and $0\le j<n$, let $\jO\defeq n-1-j$. If $\pi_j^{(n)}$ and $\piT_j^{(n)}$ are the outputs of Algorithms~\ref{natural_encode} and~\ref{214627_21Aug24}, respectively, then $\piT_j^{(n)}=\pi_j^{(n)}$.
\end{teiri}
\begin{IEEEproof}
For $0\le r<n$, let $P_r^{(n)}:=\{\pi_0^{(n)},\ldots,\pi_r^{(n)}\}$ and set $P_{-1}^{(n)}:=\emptyset$. We prove the claim by induction on $j$, simultaneously for every length $n>j$. For $j=0$, the final extension places $s^{(n-1)}$ at the first coordinate. Algorithm~\ref{214627_21Aug24} likewise selects the element of zero-based rank $s^{(n-1)}$ from $[n]$, namely $s^{(n-1)}$.

Assume the claim at coordinate $j-1$ for every applicable length, and fix $n>j$. Since $(n-1)-1-(j-1)=\jO$, the induction hypothesis applied to the length-$(n-1)$ encoder gives
\[
\pi_{j-1}^{(n-1)}
=\operatorname{select}_{s^{(\jO)}}\bigl([n-1]\setminus P_{j-2}^{(n-1)}\bigr).
\]
The defining recursion and Lemma~\ref{181528_2Aug23} therefore yield
\begin{align*}
\pi_j^{(n)}
&=\phi_{s^{(n-1)}}\bigl(\pi_{j-1}^{(n-1)}\bigr)\\
&=\operatorname{select}_{s^{(\jO)}}
\phi_{s^{(n-1)}}'\bigl([n-1]\setminus P_{j-2}^{(n-1)}\bigr).
\end{align*}
Moreover, the first $j$ entries of $\piU^{(n)}$ consist of the new head and the images of the first $j-1$ entries of $\piU^{(n-1)}$, so
\[
P_{j-1}^{(n)}=\phi_{s^{(n-1)}}\bigl(P_{j-2}^{(n-1)}\bigr).
\]
Lemma~\ref{224537_2Aug23} now gives
\[
\phi_{s^{(n-1)}}'\bigl([n-1]\setminus P_{j-2}^{(n-1)}\bigr)
=[n]\setminus P_{j-1}^{(n)}.
\]
Hence $\pi_j^{(n)}$ is exactly the element selected at step $j$ of Algorithm~\ref{214627_21Aug24}, completing the induction.
\end{IEEEproof}

At step $j$, Algorithm~\ref{214627_21Aug24} selects an element by rank from a dynamically shrinking set. An order-statistics tree supports each selection and deletion in $O(\log n)$ time~\cite{cormen}, giving total complexity $O(n\log n)$.
\begin{figure}[!t]
    \begin{algorithm}[H]
        \caption{Sequential Encoder for $\Cc^{(n)}$}
        \label{002543_29Jun24}
        \label{214627_21Aug24}
        \begin{algorithmic}[1]
        \renewcommand{\algorithmicrequire}{\textbf{Input:}}
        \renewcommand{\algorithmicensure}{\textbf{Output:}}
        \REQUIRE $ (s^{(0)} , \dotsc, s^{(n-1)})\in S^{(0)}\times\cdots\times S^{(n-1)}$\\
        \ENSURE   $(\piT^{(n)}_0,\ldots,\piT^{(n)}_{n-1})\in \Cc^{(n)}$
         \FOR {$j = 0$ to $n-1$}
         \STATE $\piT^{(n)}_j:=\operatorname{select}_{s^{(\jO)}}([n]\setminus \{\piT_0^{(n)},\ldots,\piT_{j-1}^{(n)}\})$\label{173310_26Aug24}
         \ENDFOR
         \RETURN $(\piT^{(n)}_0,\ldots,\piT^{(n)}_{n-1})$
        \end{algorithmic}
    \end{algorithm}
\end{figure}

\subsection{Bounded-Distance Decoding for REP Codes}\label{143813_28Aug23}
Encoding identifies, at each coordinate, one candidate value for every allowed head. If the heads at that stage differ pairwise by at least $d$, the corresponding candidates are also separated by at least $d$. The decoder can therefore recover the head sequence from left to right by choosing the candidate nearest to each received value.

Consider an REP code
\[
\Cc^{(n)}=\<S^{(0)},\ldots,S^{(n-1)}\>
\]
whose head sets satisfy
\[
|s-t|\ge d
\quad\text{for distinct }s,t\in S^{(j)}\text{ and every }j.
\]
Let $\sU$, $\piU$, $\rhoU$, and $\hat{\sU}$ denote the message, transmitted codeword, received array, and estimated message, respectively, where $\rhoU\in\mathbb{R}^n$, and write $\iO:=n-1-i$. At step $i$, let $\widehat P_i:=\{\piH_0^{(n)},\ldots,\piH_{i-1}^{(n)}\}$ be the set of previously reconstructed symbols, with $\widehat P_0:=\emptyset$. For $A\subset[n]$ with $|A|=i$ and $s\in[n-i]$, define
\begin{align}
\psi_i(s;A)
\defeq \operatorname{select}_s\bigl([n]\setminus A\bigr). 
\end{align}
Here $\psi_i(s;A)$ is the candidate symbol of rank $s$ after the $i$ symbols in $A$ have already been removed.
At step $i$, Algorithm~\ref{160732_26Aug24} chooses, among the candidates corresponding to the elements of $S^{(\iO)}$, one that is closest to $\rho_i$. Ties outside the guaranteed decoding radius may be resolved by any fixed rule.

\begin{figure}[H]
    \begin{algorithm}[H]
        \caption{Sequential Decoder for $\Cc^{(n)}$}
        \label{160732_26Aug24}
        \begin{algorithmic}[1]
        \renewcommand{\algorithmicrequire}{\textbf{Input:}}
        \renewcommand{\algorithmicensure}{\textbf{Output:}}
        \REQUIRE Received array $(\rho^{(n)}_0,\ldots,\rho^{(n)}_{n-1})\in\mathbb{R}^n$
        \ENSURE  Estimated message $(\sH^{(0)}\in S^{(0)},\ldots,\sH^{(n-1)}\in S^{(n-1)})$
         \FOR {$i = 0$ to $n-1$}
		 \STATE $\displaystyle\sH^{(\iO)}:=\argmin_{s\in S^{(\iO)}}|\rho_i^{(n)}-\psi_i(s;\widehat P_i)|$
         \STATE $\piH^{(n)}_i:=\psi_i(\sH^{(\iO)};\widehat P_i)$
         \ENDFOR
         \RETURN $({\sH^{(0)},\ldots,\sH^{(n-1)}})$
        \end{algorithmic}
    \end{algorithm}
\end{figure}

The decoding guarantee is the usual half-minimum-distance statement in coordinatewise form: an error smaller than $d/2$ at every coordinate cannot move the received value closer to a different candidate.
\begin{teiri}\label{221904_14Jun24}
For an integer $d\ge1$ and an REP code with $\dmin(S^{(j)})\ge d$ for every $j$, Algorithm~\ref{160732_26Aug24} returns $\hat{\sU}=\sU$ whenever $|\pi_i-\rho_i|<d/2$ for every $i\in[n]$.
\end{teiri}
\begin{IEEEproof}
We use induction on $i$. Assume that all preceding message symbols have been recovered correctly; for $i=0$ this assumption is vacuous. The reconstructed prefix is then correct, so Algorithm~\ref{214627_21Aug24} gives
$\pi_i=\psi_i(s^{(\iO)};P_i)=\psi_i(s^{(\iO)};\widehat P_i)$,
where $P_i:=\{\pi_0^{(n)},\ldots,\pi_{i-1}^{(n)}\}=\widehat P_i$.

Suppose, for contradiction, that the decoder chooses $\hat{s}^{(\iO)}\ne s^{(\iO)}$. Since both values belong to $S^{(\iO)}$ and $\dmin(S^{(\iO)})\ge d$, their difference is at least $d$. The unused symbols form a strictly increasing integer sequence, so order statistics whose ranks differ by at least $d$ also differ in value by at least $d$. Therefore,
\begin{align}
&|\psi_i(\hat{s}^{(\iO)};\widehat P_i)-
\psi_i(s^{(\iO)};\widehat P_i)|\ge d. \label{decoder_separation}
\end{align}
On the other hand, the nearest-candidate rule and the channel assumption imply
\begin{align*}
&|\psi_i(\hat{s}^{(\iO)};\widehat P_i)-
\psi_i(s^{(\iO)};\widehat P_i)|\\
&\quad\le |\psi_i(\hat{s}^{(\iO)};\widehat P_i)-\rho_i|
+|\rho_i-\pi_i|\\
&\quad\le 2|\rho_i-\pi_i|<d,
\end{align*}
contradicting \eqref{decoder_separation}. Thus $\hat{s}^{(\iO)}=s^{(\iO)}$, completing the induction.
\end{IEEEproof}

The coordinatewise hypothesis of Theorem~\ref{221904_14Jun24} is equivalent to $\dinf(\piU,\rhoU)<d/2$. Thus Algorithm~\ref{160732_26Aug24} is a bounded-distance decoder for these REP codes.

Assume that each fixed head set is stored in increasing order with random access and that an order-statistics tree stores the unused symbols. At each step, the tree evaluates $\psi_i$ in $O(\log n)$ time~\cite{cormen}. Because $\psi_i$ is monotone in $s$, binary search over $S^{(\iO)}$ finds a nearest candidate using $O(\log|S^{(\iO)}|)$ evaluations. The reconstructed symbol is then deleted from the tree in $O(\log n)$ time. Since $|S^{(\iO)}|\le n$, the overall per-codeword complexity is $O(n\log^2 n)$.
\section{Conclusions and Future Work}
The apparent simplicity of REP-code size growth hides a delayed tradeoff: a large head set can increase cardinality immediately while forcing later singleton-head extensions to restore distance. By measuring the remaining singleton-step requirement and relating it to the spacing shortfall of each head set, we converted this tradeoff into a global cardinality bound. For $n>d\ge1$, the bound is attained by equally spaced heads and equals the cardinality of the corresponding DPGP construction. The result determines the exact maximum within the REP family. The length-$4$, distance-$2$ example shows that equality of cardinalities does not imply preservation of all pairwise Chebyshev distances.

The recursive representation also yields a sequential encoder that outputs permutation symbols in coordinate order. Treating the reversed head sequence as Lehmer-code digits subject to $s^{(j)}\in S^{(j)}$ gives an $O(n\log n)$ implementation. If every two distinct elements of each head set differ by at least $d$, the sequential decoder corrects every received array at Chebyshev distance less than $d/2$ and runs in $O(n\log^2 n)$ time.

Further directions include systematic REP encoders and soft-decision, list, and error-erasure decoders. Extending the method to other permutation metrics will require new distance analyses because the present extension lemmas are specific to the Chebyshev metric.


\ifCLASSOPTIONcaptionsoff
  \newpage
\fi

\bibliographystyle{IEEEtran}
\newpage
\bibliography{IEEEabrv,../../../literature/00kasai}






\end{document}